\documentclass[aps,prl, twocolumn]{revtex4-1}
\usepackage{amssymb}
\usepackage{amsmath}
\usepackage{graphicx}
\usepackage{extarrows}
\usepackage{url}
\usepackage{varioref}
\usepackage{hyperref}
\usepackage{cleveref}
\usepackage{soul}
\usepackage{appendix}

\newcommand{\bra}[1]{\left \langle #1 \right |}
\newcommand{\ket}[1]{\left | #1 \right \rangle}

\usepackage{xcolor}

\begin{document}
%\title{EIT cooling in tripod system}
%\title{Ground-state Cooling of Large Ion Chains with an EIT Tripod Scheme}
\title{Efficient ground-state cooling of large trapped-ion chains with an EIT tripod scheme}

\author{L. Feng$^{1}$ }
\email{leifeng@umd.edu}
%\author{ W. L. Tan$^{1}$, A. De$^{1}$, A. Menon$^{1}$, A. Chu$^{1}$, G. Pagano$^{1,2}$, C. Monroe$^{1}$}
\author{W. L. Tan$^{1}$}
\author{A. De$^{1}$}
\author{A. Menon$^{1}$}
\author{A. Chu$^{1}$}
\author{G. Pagano$^{1,2}$}
\author{C. Monroe$^{1}$}
\affiliation{$^{1}$Joint Quantum Institute, Center for Quantum Information and Computer Science, and Department of Physics, University of Maryland, College Park, MD 20742 \\$^{2}$Department of Physics and Astronomy, Rice University, 6100 Main Street, Houston, TX 77005}

%\\
%\normalsize{$^{1}$Joint Quantum Institute and Joint Center for Quantum Information and %Computer Science, }\\
%\normalsize{and Department of Physics, University of Maryland, College Park, MD %20742}\\
%\\

\date{\today}

\begin{abstract}
We report the electromagnetically-induced-transparency (EIT)
cooling of a large trapped $^{171}$Yb$^+$ ion chain to the quantum ground state.
Unlike conventional EIT cooling, we engage a four-level tripod structure and achieve fast sub-Doppler cooling over all motional modes. 
We observe simultaneous ground-state cooling across the complete transverse mode spectrum of up to $40$ ions, occupying a bandwidth of over $3$ MHz.  The cooling time is observed to be less than $300\,\mu$s, independent of the number of ions. Such efficient cooling across the entire spectrum is essential for high-fidelity quantum operations using trapped ion crystals for quantum simulators or quantum computers.
\end{abstract}
\maketitle

\begin{figure*}[t!]
\begin{center}
\includegraphics[width=172mm]{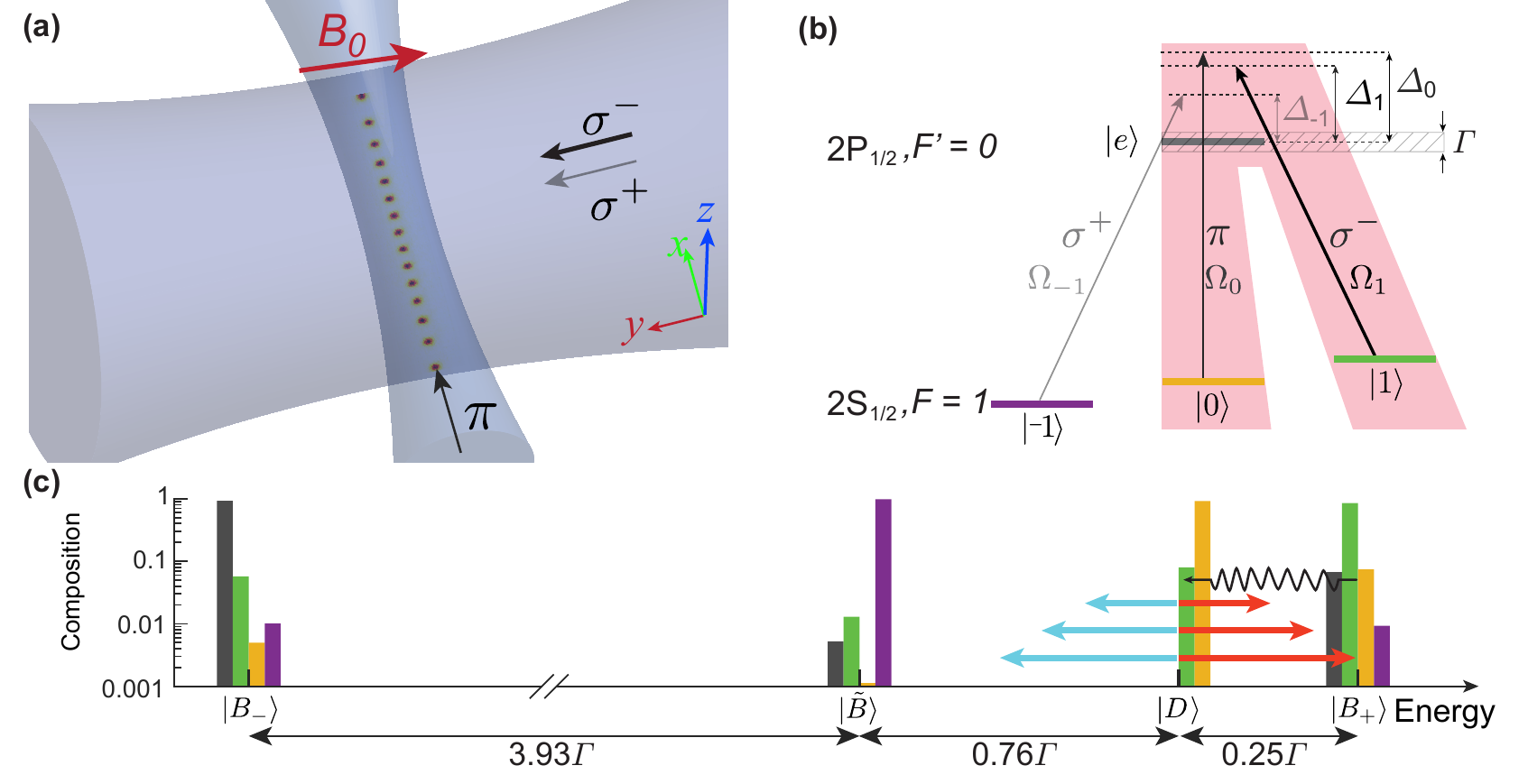}
\caption{
(a) Configuration of the EIT cooling lasers applied to a chain of trapped ions. The $\sigma^\pm$ pump beam propagates perpendicular to the chain, while the $\pi$ probe beam propagates along the chain axis ($x$). (b) Bare atomic energy levels of the tripod $^{171}$Yb$^+$ structure with coupling lasers for the EIT cooling, where the pink shading illustrates the effective EIT $\Lambda$ system. (c) Eigenstate compositions in the dressed-state energy levels, with the color code indicating the composition of the dressed eigenstates in terms of bare atomic states ($\ket{0}$: orange,$\ket{1}$: green, $\ket{-1}$: purple, and $\ket{e}$: gray).
The blue and red arrows indicate uppler and lower sideband transitions driven by motion of the ions. The black wavy arrow indicates spontaneous decay. The Rabi frequencies are $\Omega_1$~=~2.0$\Gamma$, $\Omega_0$~=~0.35$\Gamma$ and $\Omega_{-1}$~=~0.7$\Gamma$. The detunings are set to $\Delta_{0}=\Delta_{1}$~=~4.47$\Gamma$, and $\Delta_{-1}$~=~3.69$\Gamma$.
} \label{fig1}
\end{center}
\end{figure*}

\begin{figure}
    \centering
    \includegraphics[width=86mm]{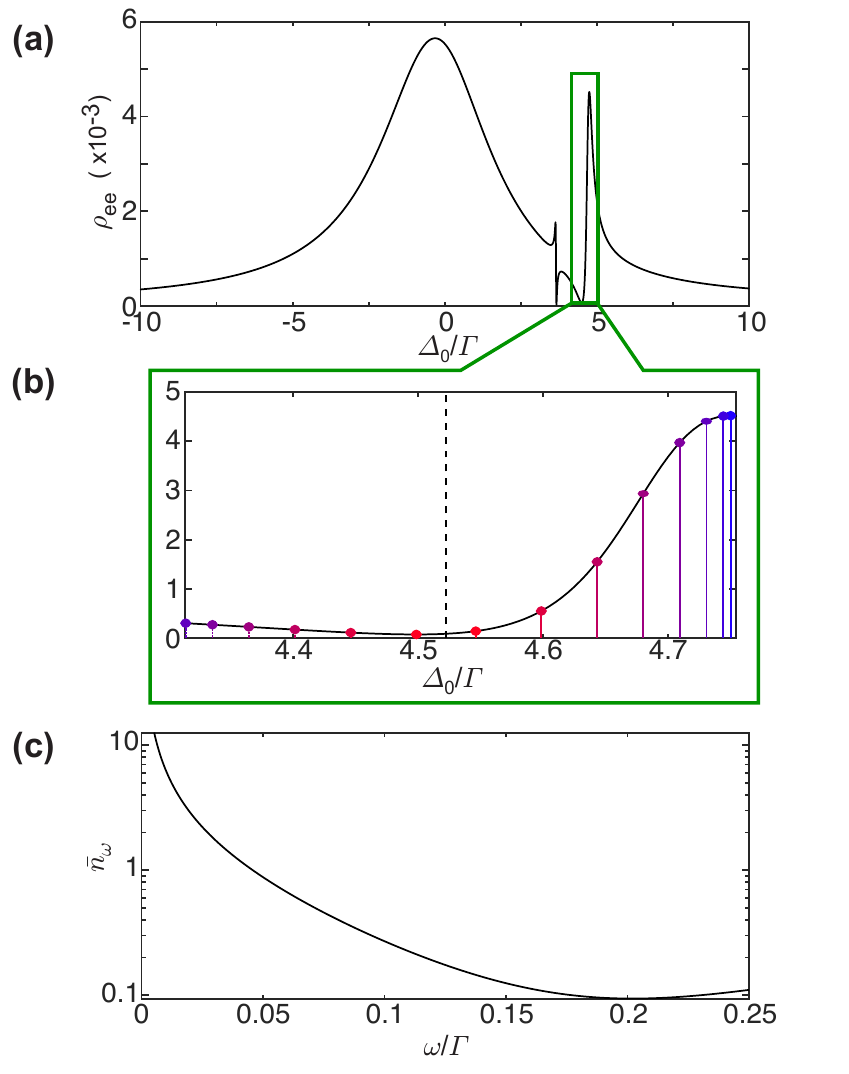}
    \caption{(a) The Fano-like absorption profile of EIT, expressed as the excited state population $\rho_{ee}$ as a function of probe laser detuning $\Delta_0$, calculated by numerically solving the steady-state solution of the master equation (see Supplement) with the same parameters as that in Fig.~\ref{fig1}c. (b)The expanded view of the EIT profile at the dark resonance of interest. The black dashed line marks the probe frequency during cooling and the red to blue dashed lines mark several example transverse mode frequencies.
%The orange dashed lines surround the dark resonance of interest. 
(c) Calculated steady-state average phonon number of motional mode with frequency $\omega$ based on EIT theory (see Supplement).}
    \label{fig1P}
\end{figure}

The laser cooling of mechanical oscillators to their motional ground state is an ongoing pursuit in quantum metrology, simulation, and computation \cite{Monroe2003,Vuletic2019, Kippenberg2008,Harold1999}.
% NEED VERY GENERAL REFS HERE!!!!  METROLOGY??
In particular, the localization of individual atoms to well below optical wavelengths (the ``Lamb-Dicke" regime) is a prerequisite for high fidelity quantum control of atomic systems \cite{Monroe2003,Saffman:2019}.
In large trapped-ion crystals, quantum entangling gates exploit the collective motion of the ions \cite{Wineland:2008, Monroe:2013}.  This motion must be prepared near the ground state in a cooling process that competes with heating from the coupling to the environment \cite{Turchette2000, Brownnutt2015}. It is therefore critical to develop new methods to achieve high-bandwidth and fast cooling of all the motional modes used as a quantum bus for quantum information processing.

Resolved sideband cooling (RSC) is a general tool for cooling mechanical oscillators, and for trapped ions it is the standard method for cooling to the ground state \cite{Wineland1989,Monroe1995,King1998,Monroe2003}. However, the RSC time typically grows linearly with the total mass of the oscillator, or the number of trapped ions in the chain. This scaling can be improved for large chains by implementing a parallel RSC strategy with single-ion addressing \cite{Chen2020}. 

Electromagnetically induced transparency (EIT) cooling of trapped ions and atoms is another well-known ground-state cooling method \cite{Keitel2000,Blatt2000, Jorg2009, Giovanna2014, Li2014, Kuhr2015, MorigiPRA2003}. It exploits quantum interference in a three-level $\Lambda$ system \cite{Tannouji1992} to create a tunable narrow spectroscopic feature tailored to the atomic motion for efficient cooling. Applied to trapped ions, EIT cooling allows simultaneous ground state cooling over a large portion of the motional spectrum without the need for single-ion addressing \cite{Bollinger2019, Holland2019, Roos2016}.
Extensions of EIT cooling beyond the simple three-level system has stimulated several theoretical \cite{Semerikov2018, Guo2015, Evers2004} and experimental \cite{Schmidt2018,Ejtemaee2017,Qiao2020} studies. 
This extension is important for quantum information applications 
with trapped-ion hyperfine qubits \cite{Blinov:2004, Monroe:2013}, whose atoms feature four or more atomic ground states.
Here, we demonstrate EIT cooling with a four-level tripod structure in a chain of up to $40$ $^{171}$Yb$^+$ ions.  
We achieve fast ground-state cooling of nearly all motional modes of the chain, occupying a broadband spectrum of more than 3~MHz, in a time ($<300\, \mu$s) that is independent of the number of ions.

The EIT cooling of the tripod $^{171}$Yb$^+$ system is implemented on the $^{2}S_{1/2}\ket{F=1} \leftrightarrow~ ^2P_{1/2}\ket{F'= 0}\equiv\ket{e}$ transition at an optical wavelength of $369.5$ nm and having a natural linewidth of $\Gamma$~=~$2\pi\times$19.6~MHz. A constant magnetic field $B_0 = 5.5$ G applied along the $y$-axis (Fig.~\ref{fig1}a) provides a Zeeman shift of $\pm\Delta_B/2\pi=\pm$7.7~MHz for the $\ket{\pm 1}\equiv\ket{F = 1, m_F =\pm 1}$ states with respect to the $\ket{0}\equiv\ket{F = 1, m_F = 0}$ state (Fig.~\ref{fig1}b). The EIT laser configuration involves two beams simultaneously and globally addressing the ions, with three components of polarization, as depicted in Fig.~\ref{fig1}b.
The beam perpendicular to the ion chain is the strong pump beam, with a large component of $\sigma^-$ and a small component of $\sigma^+$ polarization, with the power ratio controlled by a birefringent waveplate. The beam along the chain of ions is the weak probe beam with $\pi$ polarization. 

This tripod level configuration can be reduced to an effective $\Lambda$ system by setting the frequency of the $\sigma^-$ (pump) and the $\pi$ (probe) components near two-photon resonance. The $\sigma^+$ component is derived from the same laser as of $\sigma^-$, therefore it is naturally detuned from any two-photon resonance because of the Zeeman shift, but serves to remove population from the $\ket{-1}$ state by off-resonant scattering (Fig.~\ref{fig1}b).

To understand this simplification, we consider the tripod system interacting with the EIT lasers (Fig.~\ref{fig1}b), where $\Omega_{i}$ is the Rabi frequency of the laser beam that couples the ground state $\ket{i}$ ($i = \pm1,0$) to the excited state $\ket{e}$ with detuning of $\Delta_i$. 
We set $\Delta_0=\Delta_1\equiv\Delta$ and $\Delta_{-1} = \Delta-2\Delta_B$ then obtain the eigenstates.
A singular dark eigenstate $\ket{D}$ consists of the $\ket{0}$ and $\ket{1}$ atomic states while two bright eigenstates $\ket{B_\pm}$ contain a significant fraction of the excited state $\ket{e}$ along with a negligible fraction of $\ket{-1}$ (Fig.~\ref{fig1}b). The corresponding eigenvalues are $E_D \approx \Delta$ and $E_{B_\pm} \approx \frac{1}{2}(\Delta\pm\sqrt{\Delta^2+\Omega_0^2+\Omega_1^2})$, where we use $\hbar=1$.
The fourth remaining eigenstate $|\tilde{B}\rangle$ is another bright state consisting mostly of the $\ket{-1}$ state with a tiny fraction of the excited state $\ket{e}$, and largely decoupled from the bare atomic states $\ket{0}$ and $\ket{1}$. Thus we can ignore the $\ket{-1}$ state and 
consider just the effective $\Lambda$ system comprising $\ket{e}$,$\ket{0}$ and $\ket{1}$ (Fig.~\ref{fig1}c)(see Supplement).

When the EIT beams are applied to stationary ions, the steady state population is
trapped in the dark state irrespective of its initial state \cite{Tannouji1992}. However, the ion motion can drive the population out of the dark state.
When we set $E_{B_+}-E_D$ close to the oscillation frequency, the ions lose a quantum of vibrational energy as they are flipped from the dark state $\ket{D}$ to the bright state $\ket{B_+}$. The population in the bright state subsequently spontaneously decays back to the dark state $\ket{D}$.  In the Lamb-Dicke regime, the ions recoil with very low probability \cite{Monroe2003}, so this EIT absorption and emission process completes the cooling cycle, similar to RSC.

We estimate the limits of EIT cooling by considering the above cooling process balanced with heating arising from off-resonant scattering through the upper motional sidebands of the $\ket{D} \leftrightarrow\ket{B_\pm}$ transitions. We calculate the EIT absorption spectrum
by using a master equation for the full tripod system (see Supplement), and express
the EIT cooling limit as the steady-state average phonon number of a motional mode at frequency $\omega$ \cite{Monroe2003,Semerikov2018}, 
\begin{equation}
\bar{n}_\omega = \frac{\rho_{ee}(\Delta_0)+\rho_{ee}(\Delta_0-\omega)}{\rho_{ee}(\Delta_0+\omega)-\rho_{ee}(\Delta_0-\omega)}. \label{avPhnum}
\end{equation}
Here, $\rho_{ee}(\Delta_0)$ and $\rho_{ee}(\Delta_0\pm\omega)$ are the calculated steady-state populations of the excited state $\ket{e}$ at the carrier and sidebands.  
In order to apply EIT to multiple modes, the idea is to
set the Rabi frequencies $\Omega_i$ and detunings $\Delta_i$ to produce a low cooling limit for a wide spectrum of modes, as shown in Fig.~\ref{fig1P}.

\begin{figure}
\begin{center}
\includegraphics[width=87mm]{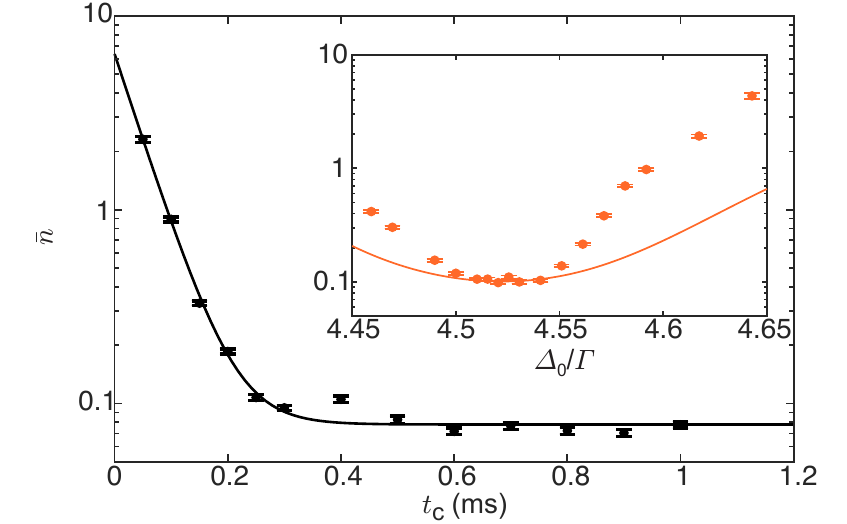}
\caption{Measured mean phonon number (black dots) as a function of EIT cooling duration $t_C$ extracted from the Raman spectroscopy for one transverse mode of the motion at $\omega_\alpha$~=~$2\pi\times$4.45~MHz for a single ion. The solid curve is a fit of the data to an exponential decay plus an offset term. 
The inset shows measurements of steady-state mean phonon number with various setting of the probe detuning $\Delta_0$, along with the master equation theory (solid line, see Supplement).
The Rabi frequencies in the experiment are $\Omega_1$~=~2.0$\Gamma$, $\Omega_0$~=~0.76$\Gamma$, and $\Omega_{-1}$~=~0.8$\Gamma$. The detunings are set to $\Delta_1$~=~4.5$\Gamma$, $\Delta_0$~=~4.54$\Gamma$ and $\Delta_{-1}$~=~3.69$\Gamma$ respectively.
We see qualitative agreement between theory and experiment, and good quantitative agreement for probe detunings used in the experiment near the optimal cooling limits.} \label{fig2}
\end{center}
\end{figure}

\begin{figure*}[t!]
\begin{center}
\includegraphics[width=172mm]{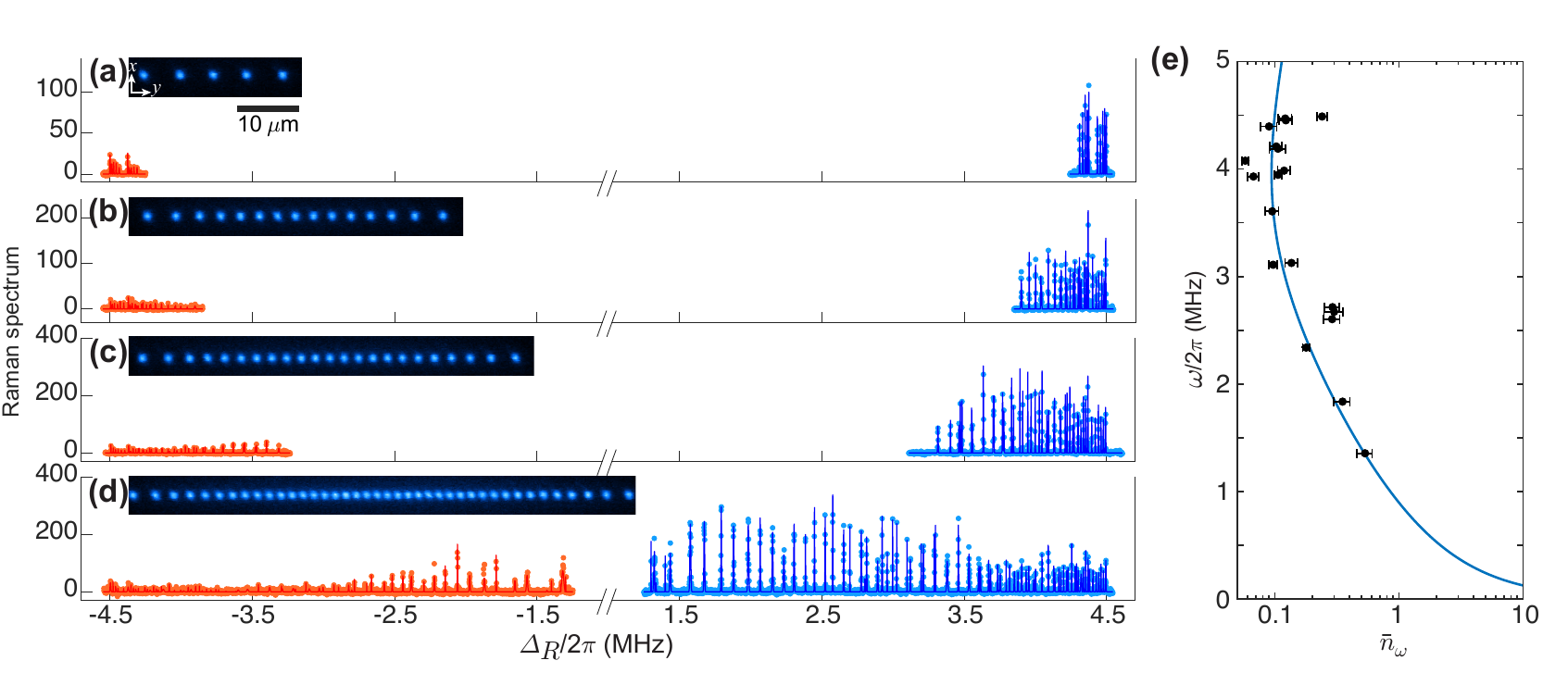}
\caption{
The lower (red) and upper (blue) motional sideband spectrum for a chain of 5, 15, 23 and 40 ions are shown in (a), (b), (c) and (d) respectively after EIT cooling. The horizontal axis is the Raman detuning from the qubit carrier transition. The highest frequency (COM) modes near 4.45 MHz do not change as the number of ions increase. With each additional ion, two more transverse modes appear at lower frequencies. The strong sideband asymmetry indicates ground-state cooling of the corresponding motional modes, over a large cooling bandwidth of 3 MHz.  The dots indicate experiment data, solid lines are Gaussian fits to guide the eye. The inset shows the ion chain  in the corresponding cooling experiment. (e) The extracted steady-state phonon number of select transverse modes across the motional band of a 40-ion chain. The dots indicate the experiment data and the solid line is the theoretical prediction.}
\label{fig3}
\end{center}
\end{figure*}

In this work, we employ a cryogenic trapped-ion apparatus \cite{Pagano2018} based on a linear Paul trap that confines the ions with transverse center-of-mass (COM) frequencies $(\omega_\alpha,\omega_\beta) = 2\pi\times (4.45, 4.30$~MHz$)$, and axial frequency $\omega_{\rm{ax}} = 2\pi \times (0.29-0.39$ kHz$)$, depending on the number of ions. 
Because both principal axes of transverse motion have components along the wavevector difference between the EIT pump and probe beams ($y$-axis in Fig. \ref{fig1}a), both directions of transverse motion are cooled by EIT.
Following this cooling, we measure the vibrational population of the transverse modes by performing conventional sideband spectroscopy.  We use a pair of counter-propagating 355~nm laser beams to drive motional-sensitive stimulated Raman transitions between qubit states $\ket{\downarrow}\equiv {^2}S_{1/2}\ket{F = 0,m_F = 0}$ and the $\ket{\uparrow} \equiv {^2}S_{1/2}\ket{F = 1, m_F = 0}$ \cite{Hayes2010}, and measure the lower/upper sideband ratio $R_\omega$ for each transverse mode and extract $\bar{n}_\omega = R_\omega/(1+R_\omega)$ \cite{Monroe2003} (see Supplement).  

We first demonstrate the EIT cooling scheme for a single $^{171}$Yb$^{+}$ ion. Beside the main EIT beams generated from a 369.5~nm diode laser,a 14.7~GHz sideband is added to the pump beam to avoid population trapping in the $\ket{\downarrow}$ state \cite{Olmschenk2007}. The Rabi frequency and laser detuning of the EIT pump and probe beams are optimized to achieve the best cooling for the transverse mode at $\omega_\alpha$ (Fig.~\ref{fig2} inset)(see Supplement).
We measure the steady-state phonon number for various cooling time $t_C$ and extract a $1/e$ cooling time of 48~$\mu s$ and the steady-state average phonon number of $\bar{n}_{\omega}$~=~0.08 with a cooling rate of 8.4$\times$10$^4$ quanta/s (Fig.~\ref{fig2}).

We next perform EIT cooling on chains of $N = 5, 15, 23$ and $40$ trapped ions following the same experimental procedure, with results shown in Fig.~\ref{fig3} (a)-(d). We see a total of $2N$ sideband features corresponding to both sets of transverse modes. We apply the EIT cooling for a fixed time $t_C$~=~300~$\mu s$, independent of the number of ions, and observe a strong suppression of the lower sidebands compared to the upper sidebands, indicating efficient cooling of the motional modes close to their ground states.
We observe average phonon numbers as low as $\bar{n}_\omega=~0.04~\pm~0.01$ for particular modes, and $\bar{n}_\omega<0.54$ for all modes over a 3~MHz bandwidth (Fig.~\ref{fig3}e), using the same amount of cooling time as that for a single ion.  The EIT cooling method is thus found to be independent of the number of ions or modes, assuming that the sideband spectrum of modes remains within the cooling bandwidth and there is sufficient laser power available to provide the same intensity on each ion. %The deviation of the experimental data at low frequency from the theory results from the instability of the mode frequency due to the low axial confinement, which is required to maintain long linear chains (see Supplement).

We finally investigate the cooling performance of a sequential combination of EIT cooling then RSC of select modes for a chain of 36 ions (See Supplement). We observe that this combination provides better cooling efficiency than either method individually. In the future, EIT cooling may be optimized further in the tripod system by applying a stronger $\sigma^+$ beam further detuned from the useful two-photon resonance. This will keep atoms away from the spectator bright state $|\tilde{B}\rangle$, and should result in even larger cooling bandwidths, lower phonon populations, and shorter cooling times, as predicted by theoretical models (see Supplement). Overall, the EIT cooling mechanism discussed here appears to be an excellent tool for the quantum control of large chains of atomic ions for quantum information applications.

We acknowledge early discussions with Kristi Beck, Michael Foss-Feig, and Tobias Grass. This work
is supported by the ARO MURI on Modular Quantum Systems, the DARPA DRINQS program, the DOE BES award de-sc0019449, the DOE HEP award de-sc0019380, and the Seed-Funding Program of the NSF Physics Frontier Center at JQI.

\bibliography{Refs.bib}

\clearpage

\setcounter{equation}{0}
\setcounter{page}{1}
\setcounter{figure}{0}
\makeatletter 
\renewcommand{\thefigure}{S\@arabic\c@figure}
\makeatother
%\underline{\smash{System performance}}\\
%test supplementary
\section{Supplement}

\subsection{Dressed-state picture of EIT cooling}

To understand the EIT cooling in the tripod structure, we start considering the Hamiltonian of a single ion interacting with the EIT lasers without spontaneous emission ($\hbar$~=~1),
\begin{gather}
H = \Delta_0\ket{0} \bra{0}+\Delta_1\ket{1} \bra{1}+\Delta_{-1}\ket{-1} \bra{-1}\nonumber\\
+\left(\frac{\Omega_0}{2}\ket{0} \bra{e}+\frac{\Omega_1}{2}\ket{1} \bra{e}+\frac{\Omega_{-1}}{2}\ket{-1} \bra{e}+h.c.\right).\label{eq1}
\end{gather}
%where $\Omega_{i}$ is the Rabi frequency of the laser beam that couples the ground state $\ket{i}$ to the excited state $\ket{e}$ with detuning of $\Delta_i$, and $i = $ -1, 0 or 1. 
When we set $\Delta_0=\Delta_1=\Delta$ while $\Delta_{-1} = \Delta-2\Delta_B$, only $\sigma^-$ and $\pi$ beams satisfy two-photon resonance condition. We can therefore neglect the contribution of the $\sigma^+$ beam and focus primarily on the strong dark feature (Fig. \ref{fig1P}) of the effective three-level $\Lambda$ system described by the simplified Hamiltonian
\begin{equation}
H = \Delta\ket{0} \bra{0}+\Delta\ket{1} \bra{1}+\left(\frac{\Omega_0}{2}\ket{0} \bra{e}+\frac{\Omega_1}{2}\ket{1} \bra{e}+h.c.\right).
\end{equation}
By diagonalizing this Hamiltonian, we find that the three eigenstates of the $\Lambda$ system can be expressed as
\begin{eqnarray}
\ket{D}&=& \frac{\Omega_1\ket{0}-\Omega_0\ket{1}}{\Omega}\label{state:D}\\
\ket{B_+}&=& \frac{\Omega_0\ket{0} +\Omega_1\ket{1}+(\Omega'-\Delta)\ket{e})}{\sqrt{2\Omega'(\Omega'-\Delta)}}\label{state:B+}\\
\ket{B_-}&=& \frac{-\Omega_0\ket{0}-\Omega_1\ket{1}+(\Omega'+\Delta)\ket{e}}{\sqrt{2\Omega'(\Omega'+\Delta)}}\label{state:B-}.
\end{eqnarray}
The corresponding eigenvalues are $E_D = \Delta$ and $E_{B_\pm} = (\Delta+\sqrt{\Delta^2+\Omega_0^2+\Omega_1^2})/2$, with $\Omega = \sqrt{\Omega_0^2+\Omega_1^2}$ and $\Omega' = \sqrt{\Delta^2+\Omega^2}$. By inverting Eq.~(\ref{state:D}, \ref{state:B+}, \ref{state:B-}), the bare atomic states can be expressed in terms of the dressed states, 
\begin{eqnarray}
\ket{0}&=& \frac{\Omega_1}{\Omega}\ket{D} + \frac{\Omega_0}{\sqrt{2\Omega'}}\left(\frac{\ket{B_+}}{\sqrt{\Omega'-\Delta}}-\frac{\ket{B_-}}{\sqrt{\Omega'+\Delta}}\right)\label{state:0}\\
\ket{1}&=& -\frac{\Omega_0}{\Omega}\ket{D} + \frac{\Omega_1}{\sqrt{2\Omega'}}\left(\frac{\ket{B_+}}{\sqrt{\Omega'-\Delta}}-\frac{\ket{B_-}}{\sqrt{\Omega'+\Delta}}\right)\label{state:1}\\
\ket{e}&=& \frac{\Omega}{\sqrt{2\Omega'}}\left(\frac{\ket{B_+}}{\sqrt{\Omega'+\Delta}}+\frac{\ket{B_-}}{\sqrt{\Omega'-\Delta}}\right). \label{eqn:e}
\end{eqnarray}

We next add an interaction term that couples the $\ket{1}-\ket{e}$ transition to a single mode of motion at frequency $\omega$ represented by creation and annihilation operators $a^\dag$ and $a$ and within the Lamb-Dicke regime \cite{Monroe2003},
\begin{equation}
H_I = \frac{\eta\Omega_1}{2}\ket{1} \bra{e}(ae^{-i\omega t}+a^\dagger e^{i\omega t}) + h.c.,
\end{equation}
where the Lamb-Dicke parameter is $\eta =\sqrt{\hbar k^2/2M\omega}$ with $k$ the wavevector of the pump beam and $M$ the mass of a single ion. Using Eqs.~(\ref{state:0}, \ref{state:1}, \ref{eqn:e}), we rewrite the interaction Hamiltonian (9) in the dressed-state basis and transform it into interaction picture.  By tuning the energy splitting between bright and dark states to be comparable to the motional mode frequency, $E_{B_+}-E_D = (\Omega'-\Delta)/2 \sim \omega$, we can focus exclusively on the lower sideband interaction that coupling state $\ket{D}$ to $\ket{B_+}$, as the other terms average to zero in the rotating-wave approximation.  We find the dressed state interaction Hamiltonian is

\begin{eqnarray} \label{DressedSidebands}
H_{I} &=&\frac{\eta\Omega_f}{2}\ket{D}\bra{B_+}e^{i\left(\frac{\Omega'-\Delta}{2}\right)t}(ae^{-\omega t}+a^\dagger e^{i\omega t}) + h.c. \nonumber\\
&=& \frac{\eta\Omega_f}{2}\left(\ket{D}\bra{B_+}a +\ket{B_+}\bra{D}a^\dagger\right),
%&+&\frac{\eta\Omega_f^{(DB_-)}}{2}\ket{D}\bra{B_-}e^{i(E_{B_-}-E_D)t}(ae^{-i\omega t}+a^\dagger e^{i\omega t})\nonumber\\
%&+&\frac{\eta\Omega_f^{(B_-B_+)}}{2}\ket{B_+}\bra{B_-}e^{i(E_{B_+}-E_{B_-})t}(ae^{-i\omega t}+a^\dagger e^{i\omega t})\nonumber\\
%&+&\frac{\eta\Omega_f^{(B_+B_-)}}{2}\ket{B_-}\bra{B_+}e^{i(E_{B_-}-E_{B_+}) t}(ae^{-i\omega t}+a^\dagger e^{i\omega t})\nonumber\\
%&+&h.c.,
\end{eqnarray}
where the effective Rabi frequency is
\begin{eqnarray}
\Omega_f&=&-\frac{\Omega_0\Omega_1}{\sqrt{2\Omega'(\Omega+\Delta)}}.
\end{eqnarray}
To complete the continuous EIT cooling cycle, the population in bright state $\ket{B_+}$ spontaneously decays back to the dark state $\ket{D}$. 

\begin{figure}
\begin{center}
\includegraphics[width=87mm]{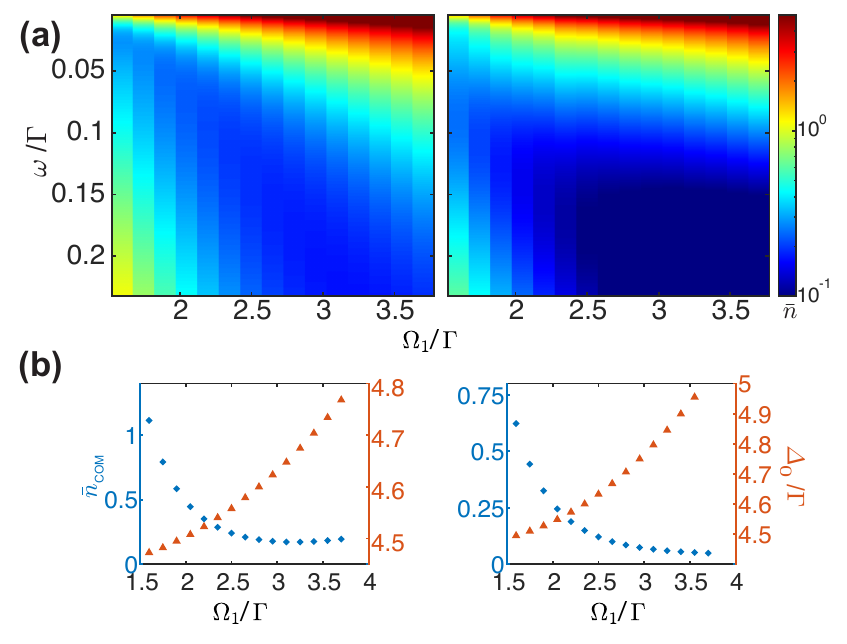}
\caption{(a) Average phonon number as a function of mode frequency and Rabi frequency of the pump beam. The detuning of the pump beam is set to $\Delta_{1}$~=~4.47$\Gamma$. Left: the detuning of the $\sigma^+$ beam is set to $\Delta_{-1}$~=~3.69$\Gamma$ and the Rabi frequency is $\Omega_{-1}$~=~0.35$\Omega_1$. Right:the detuning of the $\sigma^+$ beam is set to $\Delta_{-1}$~=~-4.47$\Gamma$ and the Rabi frequency is $\Omega_{-1}$~=~$\Omega_1$. The optimal detuning $\Delta_0$ and the corresponding average phonon number for the COM mode are show in (b).} \label{figS1}
\end{center}
\end{figure}

In order to obtain cooling limits accurately, we must include the dissipative effects of spontaneous emission along with the idealized coherent dynamics considered above.
We calculate the absorption spectrum from the steady-state solution of the master equation with the EIT Hamiltonian above,
\begin{equation}
\frac{d\rho}{dt}=-i[H,\rho]+\mathbf{L}\rho,
\end{equation}
where $\rho$ is the density matrix of the tripod system and $\mathbf{L}$ is the Lindblad operator capturing the effect of spontaneous emission, where $\mathbf{L}\rho = \sum_{j=-1}^1( b_j \rho b_j^\dagger-[ b_j^\dagger b_j,\rho ])$ with $b_j=\sqrt{\Gamma/3}\ket{j}\bra{e}$. For a steady-state solution, we take $d\rho/dt=0$.

Following the master equation, we estimate the cooling bandwidth. The excited state population for a effective $\Lambda$ system can be obtained \cite{Monroe2003},
\begin{equation}
    \rho_{ee} = \frac{4\Delta^2\Omega_0^2\Omega_1^2\Gamma}{Z},
\end{equation}
where $Z$ is given by
\begin{align}
Z = 8\Delta^2\Omega_1^2\Omega_0^2\Gamma + 2\Delta^2\Gamma^3\Omega^2\nonumber-4\Delta_0\Delta\Omega_1^4\Gamma+\frac{1}{2}\Omega^6\Gamma\\
+8\Delta^2\Gamma(\Delta_1^2\Omega_0^2+\Delta_0^2\Omega_1^2)+4\Delta_1\Delta\Omega_0^2\Gamma,
\end{align}
with $\Delta = \Delta_0-\Delta_1$. Assuming $\Delta_1\sim\Delta_0$ and $\Omega_0\ll(\Omega_1,\Delta_1)$, we have a simpler form of the excited state population 
\begin{equation}
    \rho_{ee} \approx \frac{\Delta^2\Omega_0^2}{\Delta^2\Gamma^2+4(\Omega_1^2/4-\Delta\Delta_0)^2}.
\end{equation}
As expected, the minimum appears at $\Delta = 0$ , the dark state, and the nearest maximum appears at $\Delta_0 = \frac{1}{2}(\sqrt{\Delta_1^2+\Omega_1^2}+\Delta_1)$, the bright state. 
%The half maximum locations appear at
%\begin{equation}
%    \Delta_0=\frac{1}{2}\left((\Delta_1\mp %\frac{\Gamma}{2})+\sqrt{(\Delta_1\pm\frac{\Ga%mma}{2})^2+\Omega_1^2}\right).
%\end{equation}
%Because the spectrum has a Fano-like profile, it goes to zero very slowly beyond the width at half maximum. Therefore it is not very meaningful to consider such width at half maximum of the bright state as cooling bandwidth. 
We calculate the cooling bandwidth by finding the detuning $\Delta_0$ at which the cooling stops, where the upper sideband beats the lower sideband, $\rho_{ee}(-\Delta)\ge\rho_{ee}(\Delta)-\rho_{ee}(-\Delta)$. The cooling bandwidth is thus given by 
\begin{align}
    W_C \approx \frac{1}{2}[2(1+\sqrt{2})\Delta_1+\sqrt{\Delta_
    1^2+\Omega_1^2}\nonumber\\
    -\sqrt{(3+2\sqrt(2)^2\Delta_1^2+\Omega_1^2)}]\nonumber\\
    \approx \frac{1+\sqrt{2}}{3/2+\sqrt{2}}\frac{\Omega_1^2}{\Delta_1}.
\end{align}
Given the experimental parameters, we have a cooling bandwidth of $W_C \approx 0.18\Gamma$, which is close to our observation in experiment.

Using the master equation and Eq.~(\ref{avPhnum}) in the main text, we numerically calculate the average phonon number as a function of the pump beam Rabi frequency $\Omega_1$ and trap frequency $\omega$. We are interested in cooling motional modes ranging from COM frequency 4.45~MHz ($\sim$0.23$\Gamma$) to the lowest frequency zig-zag normal mode. In the current experiment configuration, we set the detuning of weak $\sigma^+$ beam to $\Delta_{-1}$~=~3.69$\Gamma$ with the Rabi frequency $\Omega_{-1}$~=~0.35$\Omega_1$. As we increase the pump beam Rabi frequency $\Omega_1$, the lowest average phonon number improves while the cooling bandwidth reduces as well (Fig.~\ref{figS1}a). Here for each value of $\Omega_1$, we vary the detuning of the probe beam $\Delta_0$ to get the lowest phonon number for the COM mode (Fig.~\ref{figS1}b). We also find that this configuration can be further improved by increasing the Rabi frequency of the $\sigma^+$ beam while detuning its frequency further away from the two-photon resonance of interest. One case is shown with $\Delta_{-1}$~=~-4.47$\Gamma$ with the Rabi frequency $\Omega_{-1}$~=~$\Omega_1$. This also shortens the time for cooling since stronger $\sigma^+$ is more effective in keeping atoms away from $\ket{-1}$ state while not affecting the dark resonance much.

\subsection{Experimental setup for EIT cooling}

\begin{figure}
\begin{center}
\includegraphics[width=87mm]{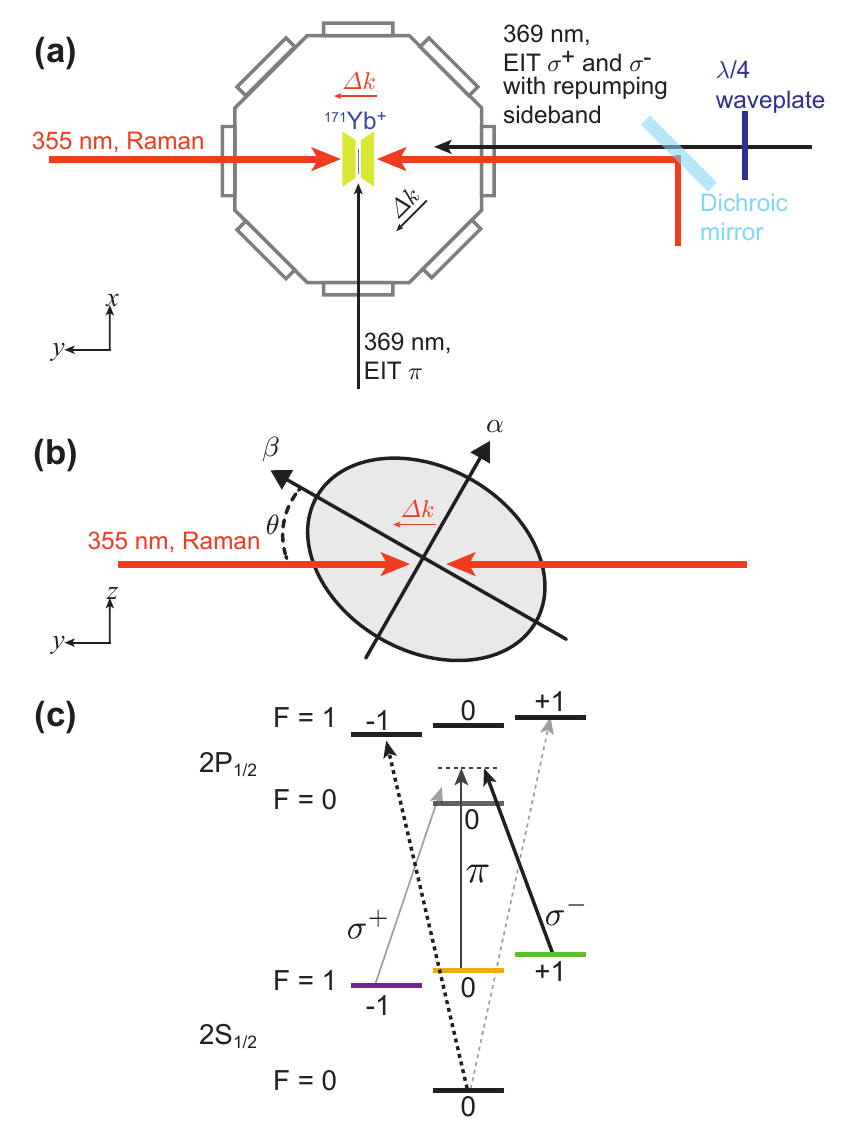}
\caption{(a) The geometric configuration of the Raman and EIT laser beams respect to ions. The $\sigma^+$ and $\sigma^-$ component in the pump laser for EIT cooling can be adjusted using the $\lambda$/4 waveplate. (b) The orientation of the Raman beams wavevector difference $\Delta k$ and the principle motional axis ($\alpha$ and $\beta$) of the ion chain. Then angle between principle axis $\beta$ and $\Delta k$ is $\theta$~=~40$^\circ$.(c) The full level diagram of $^2S_{1/2}\ket{F=1}$ to $^{2}P_{1/2}\ket{ F^{'}= 0}$ transition. In addition to the main EIT components, we also have two repumping beams (dashed lines with arrow) that are derived from the same laser using an electro-optical modulator (EOM) with a modulation frequency of 7.34~GHz and modulation depth of $\sim$0.73. The weak second-order sideband from the EOM is used for repumping atoms back to the cycling transition.} \label{figS3}
\end{center}
\end{figure}

\begin{figure*}
\begin{center}
\includegraphics[scale=1]{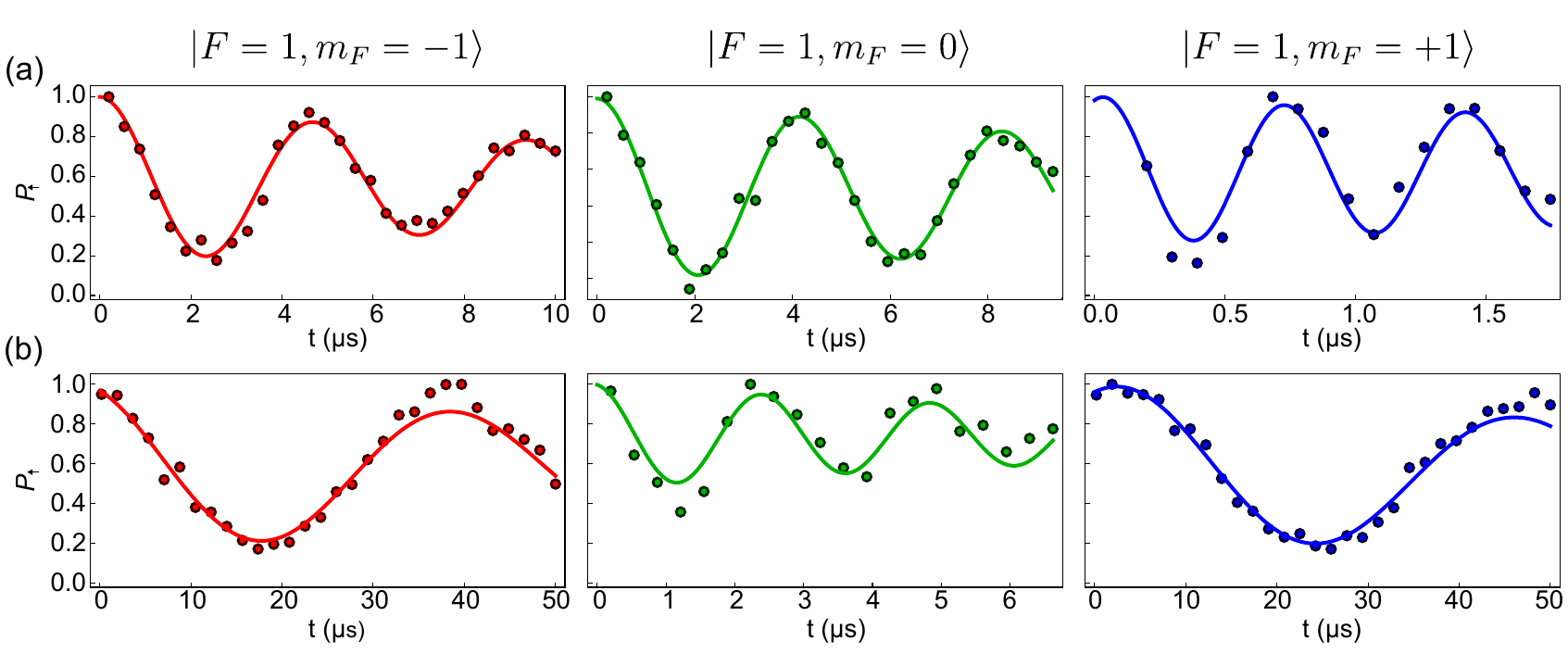}

\caption{ Left: Ramsey measurement between $^2S_{1/2}\ket{F=0,m_F=0}\leftrightarrow$ $ ^2S_{1/2}\ket{F=1,m_F=-1}$. Middle: Ramsey measurement between $^2S_{1/2}\ket{F=0,m_F=0}\leftrightarrow$ $ ^2S_{1/2}\ket{F=1,m_F=0}$. Right: Ramsey measurement between $^2S_{1/2}\ket{F=0,m_F=0}\leftrightarrow$ $ ^2S_{1/2}\ket{F=1,m_F=1}$. a) Ramsey fringes of EIT pump beam. b) Ramsey fringes of EIT probe beam.} 
\label{figLightShift}
\end{center}
\end{figure*}

The tripod-EIT scheme involves two laser beams derived from same 369.5nm laser, addressing the $^2S_{1/2}$ and $^2P_{1/2}$ manifolds (Fig. \ref{fig1}b) of $^{171}$Yb$^+$ ions. The EIT beams are configured in a way such that the wavevector difference $\Delta \vec{k} = \vec{k}_{\sigma}-\vec{k}_{\pi}$ has a component along both the transverse principle directions of motion (Fig.~\ref{figS3}b). As a result, the EIT cooling beams address all the transverse modes.
The pump beam of the EIT is along the Raman lasers (Fig. \ref{figS3}a). Two acousto-optic-modulators (AOM) in pump and probe beams are used to add different frequency shifts for each beam. The pump beam frequency is shifted by 2$\pi\times$80 MHz. As a result, the $\sigma^-$ component is detuned from the $\ket{+1} \leftrightarrow\ket{e} $ transition by $\Delta_{1}=4.5\Gamma$ and $\sigma^+$ component is detuned ($\Delta_{-1}$) from the $\ket{-1}\leftrightarrow \ket{e} $ transition by $\Delta_{-1}=3.69\Gamma$ (Fig. \ref{fig1}b). The detuning of the $\pi$ beam from the $\ket{0}\leftrightarrow \ket{e} $ transition is scanned to optimize the EIT-cooling performance for the $\omega_\alpha$ mode, and is set to $\Delta_0=4.52\Gamma$
%given by a frequency shift of 2$\pi\times$88.6 MHz from the AOM
(Fig. \ref{fig2} inset). We also employ an electro-optic-modulator (EOM) to add a 2$\pi\times$ 14.4 GHz frequency shift in the probe beam to remove the atomic population from the $\ket{\downarrow}$ state
%the cooling transition of $^2S_{1/2}$ and $^2P_{1/2}$ manifolds 
(Fig.\ref{figS2}b).

We characterize the Rabi frequencies of the corresponding optical fields using microwave Ramsey spectroscopy. We first apply a microwave $\pi/2$ pulse to prepare the ion in the superposition state of the $\ket{\downarrow}$ and $\ket{i}$ states with $i$~=~-1,0 or 1. Next, we shine each of the EIT beam separately on the atom before another microwave analysis $\pi/2$ pulse. For each EIT beam, we obtain the AC stark shift for each polarization component from the Ramsey oscillations (Fig.~\ref{figLightShift}) between the $\ket{\downarrow}$ to $\ket{i}$ with $i$~=~-1,0 or 1, which is approximately given by
\begin{equation}
    \delta_{ac}=\frac{\Omega^2}{4\Delta}.
\end{equation}\label{starkshift}
Thus from the measured stark shift, we extract the Rabi frequencies for the pump (pm) beam components,  ($\Omega_1^{pm},\Omega_0^{pm},\Omega_{-1}^{pm}$)=($2.0\Gamma, 0.8\Gamma,0.8\Gamma$). For the probe (pr) beam, we have ($\Omega_1^{pr},\Omega_0^{pr},\Omega_{-1}^{pr}$)=($0.17\Gamma, 0.76\Gamma,0.18\Gamma$). The relavent Rabi frequencies in cooling experiment are $\Omega_1\equiv\Omega_1^{pm}$, $\Omega_{-1}\equiv\Omega_{-1}^{pm}$ and $\Omega_0\equiv\Omega_0^{pr}$.
%while the rest are impurities that we have minimized \ad{(what do you mean by minimize? they are there, we can say we ignore)}.    

In EIT cooling experiment, we first calibrate all the beams and optimize the cooling for $\omega_\alpha$ mode with a single ion. Later we use the same parameters for cooling a large chain of ions. The experimental sequence involves an initial Doppler cooling of 2~ms, which reduces the phonon numbers down to $\bar{n}~\approx~5$, followed by a single pulse of EIT cooling for 300$\mu$s.

\subsection{Raman Spectroscopy}
\begin{figure*}
\begin{center}
\includegraphics[width=130mm]{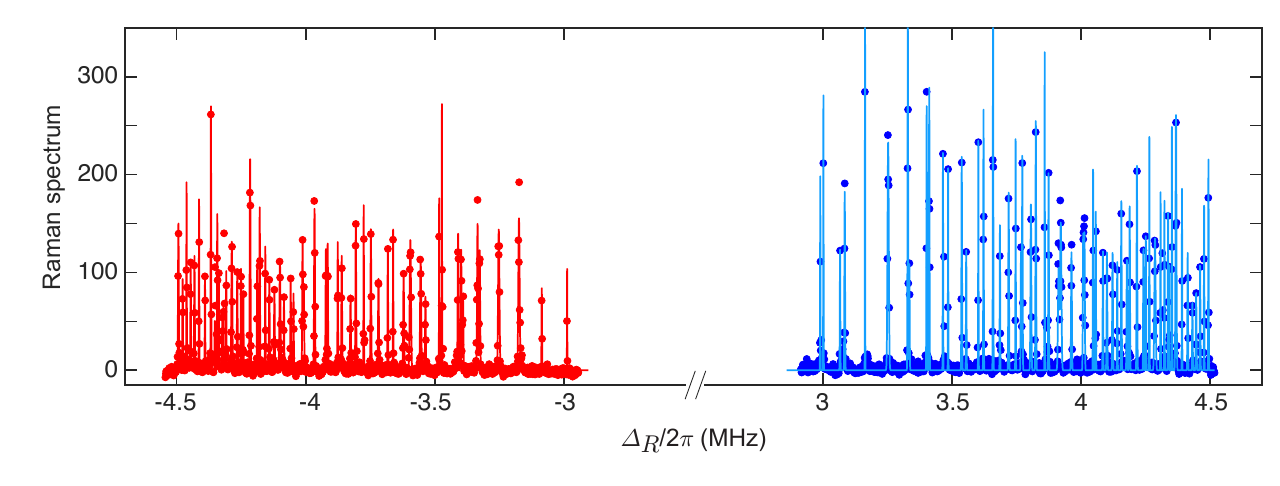}

\caption{The lower (in red) and upper (in blue) Raman sideband spectra for 31-ions chain after Doppler cooling for 2 ms. The dots indicate experimental data, solid lines are Gaussian fits to guide the eye. The motional modes for the chain have a bandwidth of 1.5 MHz. The peak heights of the red sidebands after doppler cooling can be contrasted with that of Fig. \ref{fig3}(a-d) after EIT cooling, to illustrate that EIT cooling works better than Doppler cooling over a large bandwidth.}
\label{figS2}
\end{center}
\end{figure*}

We perform conventional sideband spectroscopy on the lower/upper motional sideband of different vibrational modes. Assuming a thermal state distribution, we obtain $\bar{n}_\omega$ from the population ratio ($R_\omega$) of lower/upper sidebands following $\bar{n}_\omega = R_\omega/(1-R_\omega)$ (\cite{Monroe2003}).

When performing sideband spectroscopy with large chains, we reduce the Raman laser power until motional mode frequencies are well-resolved. In addition, we lower the axial confinement to maintian the one-dimensional chain as we add more ions to the trap. As a result, the transverse mode separation exhibits strong fluctuates between measurements. To correctly measure the population ratio ($R_\omega)$ and estimate the statistical error, we scan the motional mode frequencies 5 times for each data point in Fig.~\ref{fig3}e. Fig. \ref{figS2} shows the Raman sideband spectra for 31-ions chain after Doppler cooling for 2 ms.

\subsection{Motion-Sensitive Carrier Rabi Flopping}
As we scale up the system size, the mode spectrum becomes dense, and spectrally isolating a single motional mode to determine the thermal number of quanta in each mode becomes challenging. An alternative metric for the efficacy of EIT cooling is to measure the motion-sensitive Rabi dynamics of the ions. This provides a global measure of the thermal motion of the ions. 

For a collection of modes with phonon numbers $\{n_m\}$, the carrier Rabi frequency on a particular ion $i$ includes Debye-Waller factors that suppress the interaction.
Assuming each mode $m$ of motion is in a thermal state with mean phonon number $\bar{n}_m$, we find \cite{Monroe2003},
\begin{equation}
\bar{\Omega}_{i}
= \Omega_i\exp\left[{-\sum_m\eta_{im}^{2}(\bar{n}_m+1/2)}\right]. \label{eqn:RabiFreq}
\end{equation}
Here we take each mode to be confined within the Lamb-Dicke limit where $\eta_{im}^2(n_m+1/2) \ll 1$.
The population of a given ion initially prepared in state $\ket{\downarrow_i}$ is given by
\begin{equation}\label{SingleIon}
    P^{(i)}_{\uparrow}(t)= 
\frac{1-C_i(t) \cos(\bar{\Omega}_i t-\phi)}{2}
\end{equation}
with the phase $\phi \ll 1$.  The contrast of the Rabi oscillations is 
\begin{equation}
    C_i(t)=\prod_m\frac{1}{ \sqrt{1+(\eta_{im}^2\bar{n}_m\bar{\Omega}_i t)^2}}
    \approx1 - \frac{1}{2}(\bar{\Omega}_it)^2
    \sum_m\eta_{im}^4\bar{n}_m^2,\label{eqn:Constrast}
\end{equation}
where the last approximation assumes sufficiently early evolution times. 
We see that due to motion, the measured Rabi frequency for each ion is suppressed and the contrast decays in time. Both observables can provide a global signature of the thermal motion.

\begin{figure*}[t!]
    \centering
    \includegraphics[scale=1.0]{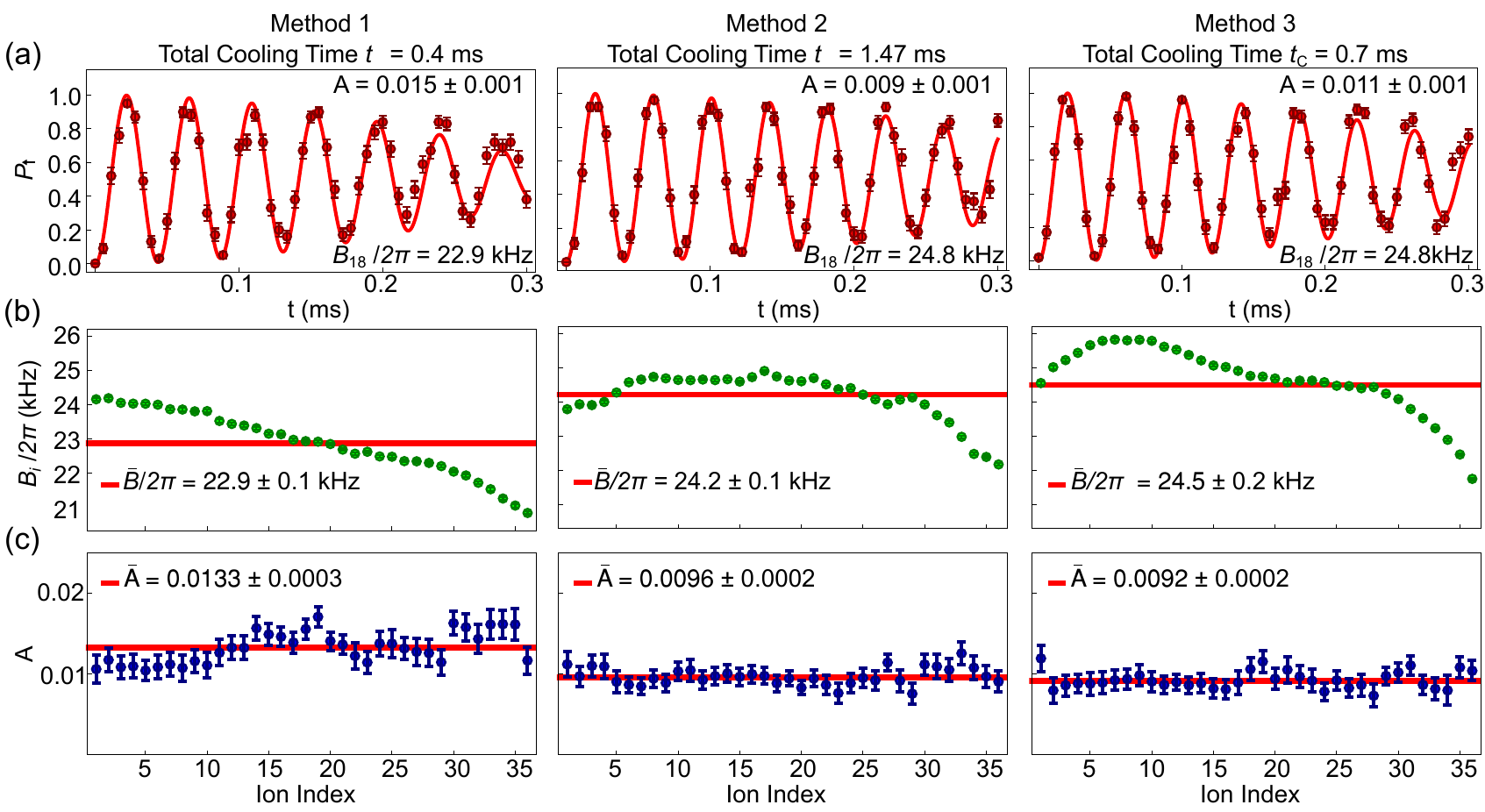}
    \caption{The performance of a 36-ion chain is characterized by motion-sensitive carrier Rabi oscillations. The transverse motional modes of the chain are spread between $2-4.45$~MHz. 
    We show (a) measured Rabi flopping of the 18th ion, (b) the measured best-fit Rabi oscillation frequency $B_i$ of each of the 36 ions, and (c)  the measured best-fit quadric contrast $A$ of each ion (see Eq. \ref{eqn:Constrast}). Each column shows the results from three different cooling schemes. Left: 40 cycles of RSC on the  two transverse COM modes at 4.45~MHz and 4.35~MHz, with a total cooling time of 0.4~ms. Middle: 40 cycles of RSC cooling with four additional frequencies at 3.57~MHz, 3.25~MHz, 2.43~MHz and 2.07~MHz with a cooling time of  1.47~ms. Right: Combination of EIT cooling (0.3 ms cooling time) and 40 cycles of RSC at 2 COM frequencies with a total cooling time of 0.7 ms. The dots in the figures indicate experimental data. The red solid curve in (a) is the fit to Eq.~\ref{RamanFitFunction}. The solid lines in (b) and (c) show average value. The errorbar indicates one standard error. The asymmetry of $B_i$ across the chain likely arises from uneven micromotion across the chain with a length of $\sim$76~$ \mu m$. 
    The fits to contrast in (c) only show a slight dependence on cooling methods, likely due to other Rabi decay processes insensitive to motion.}
    \label{fig:Raman}
\end{figure*}

We cannot extract the mean phonon number for each mode $\bar{n}_m$ based on the global observables for the Rabi frequencies (Eq. \ref{eqn:RabiFreq}) or Rabi flopping contrast (Eq. \ref{eqn:Constrast}).  However, by fitting the observed Rabi evolution of each ion, we can extract global information about the thermal motion of the ions and thereby gauge the efficacy of EIT cooling strategies.

We fit observed Rabi oscillations of each ion in a chain to  Eq.~\ref{SingleIon}, according to the fit function
%accurately to extract $\bar{n}$ for $2N$ modes is computationally difficult in longer ion chains and the Rabi oscillation of the system has an intrinsic decay due to Raman beam pointing noise, shot noise and intensity noise. This intrinsic decay of the system is the same across the same set of data since they are taken within the time frame of the slow drift of the laser beams with respect to the center of the ion chains. Therefore, we simplify Eq.\ref{SingleIon} with Eq.\ref{Constrast} and obtain the following fitting function,
\begin{equation}\label{RamanFitFunction}
    P^{(i)}_{\uparrow}(t)= \frac{1-[1-A~(B_it)^2 ]\cos(B_i t)}{2} + P_0 
\end{equation}
and extract values of the quadratic contrast decay $A$ (nominally $A = \frac{1}{2} \sum_m\eta_{im}^4\bar{n}_m^2$) and the Rabi frequencies $B_i$ for each ion.  We also allow for an offset term $P_0$ to account for detection errors. 

We use these global observables to gauge three different cooling schemes which we label method 1, 2, and 3 (Fig.~\ref{fig:Raman}). With method 1, we implement RSC on only two COM modes, thus cooling only these and neighboring modes. With method 2, we apply additional RSC at 4 more frequencies covering the full motional spectrum to ensure all the modes are cooled. With method 3, we combine broadband EIT cooling and RSC in method 1. Comparing the results with methods 1 and 3, we find that EIT cooling over a large bandwidth increases the Rabi frequency, indicating improved global cooling of the chain. With method 2, we see that RSC with full bandwidth coverage improves the Rabi flopping across the chain, which is comparable to that with method 3. However, it costs twice as much time as that of the more efficient method 3.
We find that the fitted contrast improves along with the Rabi frequency for methods 2 and 3, although only weakly, likely due to other decay mechanisms insensitive to motion such as intensity noise. 

\end{document}